\documentstyle[aaspp4]{article}

\lefthead{Driver et al.}
\righthead{Morphological number-counts and redshift distributions}

\begin{document}

%%\tighten

\title{Morphological number-counts and redshift distributions to $I < 26$ from
the Hubble Deep Field: Implications for the evolution of Ellipticals, Spirals
and Irregulars}

\author{S.P.Driver, A. Fern\'andez-Soto, W.J. Couch}
\affil{School of Physics, University of New South Wales
Sydney, NSW 2052, Australia} 

\author{S.C.Odewahn}
\affil{Department of Astronomy, Californian Institute of Technology,
Pasadena, CA 91125, USA}

\author{R.A.Windhorst}
\affil{Department of Physics and Astronomy, Arizona State University,
Box 871504, Tempe, AZ 85287-1504, USA}

\author{S. Phillipps}
\affil{Astrophysics Group, Dept. of Physics, University of Bristol,
Tyndall Avenue, Bristol, BS8 1TL, UK}

\author{K. Lanzetta, A. Yahil}
\affil{Department of Physics and Astronomy, State University of New York at 
Stony-Brook, Stony Brook, NY 11794-2100, USA}

\begin{abstract}
We combine the photometric redshift data of Fern\'andez-Soto {\it et al.} 
(1997)
with the morphological data of Odewahn {\it et al.} (1996) for all galaxies
with $I < 26.0$ detected in the Hubble Deep Field. From this combined catalog
we generate the morphological 
galaxy number-counts and corresponding redshift distributions
and compare these to the predictions of {\it high normalization} zero- and 
passive- evolution models. From this comparison we conclude the following:

\noindent
(1) E/S0s are seen in numbers and over a redshift range consistent with zero-
or minimal passive- evolution to $I = 24$. Beyond this limit fewer E/S0s
are observed than predicted implying a net negative evolutionary process 
--- luminosity dimming, disassembly or masking by dust --- at $I > 24$.
The breadth of the redshift distribution at faint magnitudes implies strong
clustering or an extended epoch of formation commencing at $z > 3$.

\noindent
(2) Spiral galaxies are present in numbers 
consistent with zero-evolution predictions to $I = 22$. 
Beyond this magnitude some net-positive evolution is required. Although the 
number-counts are consistent with the passive-evolution predictions to $I=26.0$
the redshift distributions favor number and luminosity evolution although few 
obvious mergers are seen (possibly classified as Irregulars). We note that 
beyond $z \sim 2$ very few ordered spirals are seen suggesting a formation 
epoch of spiral galaxies at $z \sim 1.5$ -- $2$. 

\noindent
(3) There is no obvious explanation for the late-type/irregular class and this
category requires further subdivision. While a small fraction of the population
lies at low redshift (i.e. true irregulars), the majority lie at redshifts, 
$1 < z < 3$.  At $z > 1.5$ mergers are frequent and, taken in conjunction 
with the absence of normal spirals at $z > 2$, the logical inference is that 
they represent the progenitors of normal spirals forming via hierarchical 
merging.

\end{abstract}


%%\twocolumn

\section{Introduction}
The Hubble Deep Field (HDF; Williams {\it et al.} 1996) has provided the 
deepest and clearest window to date on the extragalactic sky. From this 
dataset, groups have studied the morphologies of the faintest galaxies
(e.g. Odewahn {\it et al.} 1996; Abraham {\it et al.} 1996) and made 
photometric estimates of the redshift of these objects (e.g. Lanzetta, Yahil \&
Fern\'andez-Soto 1996;
Brunner {\it et al.} 1997). Here we combine these two independent analyses 
to generate morphological number-counts {\it and} morphological redshift 
distributions for a complete sample of objects from the Hubble Deep Field (413 
objects to $I=26.0$). This represents a unique dataset which provides strong 
constraints on the many faint galaxy models which 
have been postulated to explain the phenomena of the faint 
blue excess (see Ellis 1997 for a recent review). 

Faint galaxy models fall into three broad generic categories: dwarf-dominated 
models; pure luminosity evolution models; and merger models. All of these 
various
models can provide a fit to the observed faint galaxy number-counts and
therefore these data alone are insufficient to distinguish between the 
proposed models. Additional observational constraints are required and the
most definitive one is that of the observed redshift distributions, N(z), for 
progressively fainter magnitude slices. For example, dwarf-dominated 
models (Driver {\it et al.} 1994; Phillipps \& Driver 1995; Babul \& Ferguson 
1996) predict an additional low redshift component at faint magnitudes when 
compared to the N(z) predictions of the zero-evolution models. Conversely 
pure-luminosity evolution models (e.g. Metcalfe {\it et al.} 1995; Campos \& 
Shanks 1997) predict
a high redshift component whilst merger models lie somewhere in between (e.g. 
Carlberg 1992; Rocca-Volmerange \& Guiderdoni 1990). In theory then, the
problem is surmountable; in practice obtaining a comprehensive and complete
spectroscopic redshift distribution at faint magnitudes is beyond our
technological capabilities. The very faint redshift surveys which do exist 
(e.g. Glazebrook {\it et al.} 1995a; Cowie {\it et al.} 1996) are relatively 
small samples and arguably susceptible to selection biases (e.g. wavelength 
coverage, spectral features, surface brightness). For the moment 
the only recourse for establishing the N(z) distribution at these faint 
magnitudes is to utilize distance estimates based on multi-band photometry, 
{\it i.e.} photometric redshifts.

In \S 2 we briefly discuss the adopted morphological and photometric 
HDF catalogs.
In \S 3, we  compare the resulting galaxy number-count data and redshift 
distributions to zero- and passive- evolution models, and 
infer the generic form of evolution implied by the data.
\S 4 summarises our findings.

\section{The catalogs}
In recent years the high-resolution imaging provided by the Wide Field and 
Planetary 
Camera 2 on the {\it Hubble Space Telescope} has opened up the new field
of the morphological classification of faint galaxies. The initial work relied 
on the eyeball consensus of
a number of experienced galaxy classifiers to sub-divide the faint galaxy
population into ellipticals (E/S0), early-type spirals (Sabc) and late-type 
spiral/irregulars (Sd/Irr), e.g. Driver, Windhorst \& Griffiths (1995); Driver 
{\it et al.} (1995); Glazebrook {\it et al.} (1995b). The primary motivation
to sub-divide into these three categories was based on the existence of 
observable structural distinctions between these broad groupings
(e.g. de Vaucouleurs, exponential or asymmetric profiles for E/S0, Sabc or 
Sd/Irr respectively). In addition the E/S0 and Sabc classes are also known
to have distinct physical properties ({\it i.e.} 
pressure or rotationally supported components). 
If these differing structural and physical properties are the result of 
independent evolutionary paths, then this provides justification for studying 
their number-density evolution independently. Ultimately automated methods are 
required to construct statically representative samples in a fully reproducible
and objective manner. This approach has generally taken two paths: Artificial 
Neural Networks (ANN, see Odewahn {\it et al.} 1996; Naim {\it et al.} 1995) 
and decision trees based on structural parameters (e.g.
Casertano {\it et al.} 1995; Abraham {\it et al.} 1996). Both techniques show 
promise, performing at a level close to that of eyeball classification but
occasionally seen to fail when dealing with objects with overlapping isophotes.
To overcome this potential bias we initially adopt the HDF catalog generated 
by the ANN of Odewahn {\it et al.} (1996) and compared it with the results
from the eyeball classifications of one of us (WJC)\footnote{Note that the 
ANN was originally trained on data classified by SCO, RAW and SPD and hence 
this represents an unbiased and independent check of the classification 
accuracy.}.
Figure 1 shows the resulting histogram distribution of $\Delta$ T-types and can
be approximated by a Gaussian of mean $0.1 \pm 0.2$ and FWHM $1.5 \pm 0.2$. 
However, the wings are non-Gaussian suggesting that where there is 
disagreement it tends to be large and systematic. 
Examining only those images for which the T-type differences are large
show that it is routinely the ANN which has failed and that the majority of
these discrepancies are unambiguous cases of extreme irregularity, mergers or
bright cores embedded in irregular halos.
This is not surprising as there were few such objects in 
the ANNs original training set (based on brighter galaxies in less deep fields
c.f. Odewahn {\it et al.} 1997).
To accommodate for this we constructed a final catalog in which the 
initial blind ANN classifications were replaced by the eyeball classification 
if the T-type disagreement was greater than 4 T-types ({\it i.e.} where the 
histogram in Fig. 1 becomes asymmetric). This resulted in 20\% of the 
original ANN HDF galaxy classifications being overridden. Also shown on Fig. 1
is the $\Delta$ T-type histogram for the faintest magnitude bin ($25 < I < 26$)
and we note that the distribution is similar to that of the entire sample,
implying little or no degradation in classification accuracy with apparent
magnitude.

The photometric redshift 
catalog (described in more detail in Fern\'andez-Soto {\it et al.} 1997) 
is based on that presented in Lanzetta, Yahil \& Fern\'andez-Soto (1996), but 
updated to take advantage of ground-based IR images of the HDF (Dickinson 
{\it et al.} 1997). The determination of the photometric redshifts are 
obtained by maximizing the Likelihood Function $L(z,T)$.
The likelihood $L(z,T)$ of obtaining measured fluxes $f_i$ with uncertainties
$\sigma_i$ given modeled fluxes $F_i(z,T)$ for a given spectral type T at
redshift $z$, with a flux normalization $A$ over the seven filters ($i=1-7$)
is:

$$ L(z,T) = \prod_{i=1}^{7} exp \left\{ - \frac{1}{2} \left[
             \frac{f_i-AF_i(z,T)}{\sigma_i} \right] ^2 \right\} $$

This formula is maximized for each of four possible spectral types for 
each object. This results in four redshift likelihood functions
$L(z)$ which are simultaneously maximized to give both the optimal redshift 
and the spectral classification. The four model templates ($F_i(z,T)$) were 
adopted from Coleman, Wu \& Weedman (1980) and extended to the 
infrared wavelengths using the models of Bruzual \& Charlot (1993). 
Intergalactic HI absorption was taken into account 
in the same way as described in Lanzetta {\it et al.} (1996).
Details of the method and the reliability of the photometric redshifts are 
discussed in full in Fern\'andez-Soto {\it et al.} (1997).
To summarize, the agreement between the photometric and the known spectroscopic
redshifts is excellent in the low-redshift range ($z<1.5$), where the $rms$ 
deviation is $\Delta(z_{spec}-z_{phot})_{rms} \approx 0.15$ with
no measurable bias in the residual distribution. At
$z>2$ some discordant values are seen (less than 10\%) and we note
that the redshifts tend to underestimate the real values in the range 
$2 < z < 3$. No trend in $\Delta$ z with T-type was seen. 
For objects where spectroscopic redshifts were available, these
were used instead of the photometric values. The final catalog is therefore
based on 20\% spectroscopic and 80\% photometric redshifts, however we
note that the agreement between spectroscopic and photometric redshifts
are extremely good (c.f. Hogg {\it et al.}) and the results are unchanged if 
purely photometric redshifts are used.

The morphological and photometric redshift catalogs
were merged by matching the x and y positions of the two catalogs.
Whilst the majority of objects were successfully matched a small fraction
($\sim 5$\%) failed. These miscreants were traced to either positional 
discrepancies or due to differences in the deblending of complex structures,
in these cases deference was once again given to the eyeballed classifications.
The final matched catalog contains 401 objects with z's and morphologies to 
$I = 26$ and a further 12 with morphologies only ({\it i.e.} a reliable
photometric redshift was not measured or not measurable).
True colour representations of the full final catalog ordered according to 
morphology (Plate 1) and redshift (Plate 2) are shown.
Of particular interest in Plate 2 is the trend towards higher 
irregularity and/or a higher merger rate beyond $z \sim 1.5$ 
(coincident with the
peak in the star-formation rate as reported in Madau {\it et al.} 1996).
Qualitatively at least this visually suggests a Universe at $z > 1.5$ 
dominated by a period of hierarchical merging of star-forming irregular galaxy 
clumps (c.f. Pascarelle {\it et al.} 1996).

\section{Morphological N(m) and N(z)s}
Figures 2 and 3 show the morphological number-counts and morphological 
redshift distributions respectively. The zero-evolution model predictions
are shown as solid lines while the passive-evolution predictions are shown
as broken lines. These models are based
on the following: the local morphological luminosity distributions of Marzke 
{\it et al.} (1994); the k- and evolutionary- corrections of Poggianti (1997); 
a standard flat cosmology ({\it i.e.} $q_{0}=0.5, \Lambda=0$); and a global 
renormalization ($\times 1.8$) of the local galaxy numbers based on an optimal 
count match at $b_{J}=18.0$ (more detail on the modeling is given in
Driver {\it et al.} 1995b).
Considering each population independently we note:
~

\noindent
{\bf E/S0s}: At brighter magnitudes ($I < 24$), both the counts and the
redshift distributions agree well with the predictions of the zero-evolution
model, implying little net evolution (see, e.g. Driver {\it et al.} 1996 for
discussion of the counteracting effects of luminosity and number evolution).
At fainter magnitudes ($24 < I < 26$) there is a marginally
significant ($2 \sigma$) shortfall in the counts compared to the models. This
may be a statistical 
fluctuation\footnote{Given the limited statistics and strong clustering 
behavior of ellipticals, a single sight-line with a small field-of-view is 
highly susceptible to 
statistical vagaries} or a pointer towards a net-negative evolutionary process 
or a higher dust content than allowed for in the models (e.g. Campos \& Shanks 
1997). If it is real, the shortfall may indicate a modest rate of continued 
formation of some extra ellipticals via mergers (see e.g. summary in Zeigler \&
Bender 1997). These conclusions from the counts appear to be in good agreement 
with those obtained from direct structural (Fundamental Plane) and 
spectroscopic (stellar population) studies of high redshift E/S0s (Kelson 
{\it et al.} 1997; Barger {\it et al.} 1997; Ellis {\it et al.} 1997). 
It is also worth noting the small
spikes of E/S0s observed at z=3 at $24 < I < 26$; could these represent the 
end of the main formation epoch for early type galaxies or the presence
of a large scale-structure ?

~

\noindent
{\bf Sabcs}: To $I=24.0$, {\it i.e.} equivalent to the deepest 
sight-lines probed by non-HDF fields, (Driver {\it et al.} 1995), no evolution
in luminosity or number is required. Beyond this limit the zero-evolution
model underpredicts the number-counts implying some form of net positive
luminosity- or number-evolution. As the Poggianti {\it et al.} passive 
evolution model slightly overpredicts the counts, one might perhaps conclude 
that the true picture simply lies between the zero- and passive- evolution 
models. However, it is very striking that in the redshift distribution at 
$24 < I < 26$ show an excess at $z=1.5$ followed by a sharp drop at $z > 1.5$.
This is inconsistent with the passive evolution model and implies strong 
number-evolution. Direct examination of the Sabc images on Plate 1, suggests 
little evidence of merging or an overly high density of close companions, 
which would be expected if numbers were not conserved.

~

\noindent
{\bf THE Sd/Irr/M+Pec}: This class, effectively the catch all, is in 
disagreement with the models at all magnitudes. 
Significant effort is therefore required to explain this population, which is
identifiable as that responsible for the Faint Blue Excess (see e.g. Ellis 
1997). Surprisingly the redshift distribution of this population 
is very broad and has a higher mean over all magnitude intervals
than the E/S0 and Sabc population ({\it i.e.} contains a more luminous 
population than giant ellipticals and spirals).
The density of low redshift objects is roughly as expected from the models, 
{\it i.e.} the true (low luminosity) irregulars are seen in the expected 
numbers for a passively evolving population of dwarfs. (Recall that we use the 
Marzke {\it et al.}'s LF which is quite steep for Sd/Irrs). The excess 
irregulars are then predominantly at $z > 1$. The wide spread in the $z$ is 
reminiscent of the predictions for star-bursting dwarf-dominated models 
(Phillipps \& Driver 1995), the ``dwarfs'' being able to reach high 
luminosities during their initial burst phase (e.g. Wyse 1985). Nevertheless,
these objects may well also include the precursors of modern day spirals as 
well as ``genuine'' dwarfs (for instance, it is easy to see that they could
``fill in'' the redshift distribution of the Sabc class at
$z > 1.5$). Examination of Plates 1 and 2 suggests that the objects 
classified as high-$z$ irregulars are frequently seen with close companions 
and/or tidal features indicative of merging, the implication is an
epoch of merger induced star-formation occurring in the redshift interval 
$z = 1.5 - 2$.

\section{Conclusions}
We have combined the morphological catalog of Odewahn {\it et al.} (1996)
with the photometric redshift catalog of Lanzetta {\it et al.} (1996)
for all objects in the Hubble Deep Field to $I = 26$. This
has resulted in a unique dataset from which we can construct the observed
number-counts {\it and} redshift distributions for E/S0s, Sabcs and 
Sd/Irrs down to $I = 26$. Adopting the local morphological luminosity functions
(Marzke {\it et al.} 1994) and with the caveat of {\it a uniform overall 
renormalization at $b_{j} = 18$}, we conclude the following:

Ellipticals form over an extended period starting at $z > 3$, however
the observed underdensity in the number-counts implies that young ellipticals
are either masked by dust or only become recognizable morphologically as 
ellipticals after their stellar population has stabilized and aged ({\it i.e.} 
a substantial population of young overly luminous ellipticals is not seen).
From the observed absence of $L_{*}$ spirals at moderate to high redshifts 
($z > 2.0$) we conclude that present-day disks are forming at $z \sim 2$ via
hierarchical merging. During this stage their morphologies are highly 
irregular, this is corroborated by the high number of irregulars seen at this
epoch. At lower z the merger rate sharply declines and the more luminous 
(massive ?) objects crystallize into the regular spiral systems and evolve 
passively with minimal further merger events. Meanwhile the remaining less
luminous disk systems and merger by-products/remnants fade ($z > 1$) into the 
local dwarf and low surface brightness populations.

Our final {\it Hubble Deep Field} catalog of morphologies and photometric z's
is available on request from: spd@edwin.phys.unsw.edu.au

\acknowledgements
We would like to thank the HDF team and STScI staff for providing the HST and
KPNO data to the community. SPD, AFS and WJC thank the Australian Research 
Council for support and SP acknowledges the Royal Society for support via a 
Royal Society University Research Fellowship. The following HST grants are also
acknowledged: GO-5985.01-95A (RAW); AR-6385.0*-95A (RAW \& SCO) 
and GO-6610.01-95A (RAW).

\section*{Figure Captions}

\figcaption{The $\Delta$ T-type histogram for the full sample (solid line)
and the faintest magnitude bin only (dashed line). \label{fig0}}

\figcaption{Number counts for the different morphological 
types: all galaxies (upper left), E/S0 (upper right),
Sabc (lower left) and Sd/Irr (lower right). Data are from
Casertano {\it et al.} (CRGINOW), Driver, Windhorst \& Griffiths 1996 (DWG),
and Driver {\it et al.} (DWOKGR) and this study (HDF).
The number count predictions of the zero- and passive- evolution 
models are shown as solid and broken lines respectively. \label{fig1}}

\figcaption{Morphological redshift distributions for $22<I<23$ 
(top), $23<I<24$ (upper middle), $24<I<25$ (lower middle) and $25<I<26$ 
(bottom).
The columns are, from left to right: all galaxies, E/S0s, Sabc's and Sd/Irrs.
Overlaid are the zero- (solid) and passive- (broken) evolution model
predictions. \label{fig2.ps}}

\section*{Plate Captions}

\figcaption{PLATE 1: The Hubble Deep Field galaxies sub-divided according to 
their morphological classifications. The galaxies are 
ordered from left to right according to apparent magnitude. Note that colour 
information is not used in the classification process and classifications were 
made in the longest waveband filter to minimise possible 
miss-classification due to UV 
irregularities (c.f. Giavalisco {\it et al.} 1996).}

\figcaption{PLATE 2: The Hubble Deep Field photometric redshift sample.
The sample is first sorted into redshift and divided into 16 redshift bins, 
each containing 25 galaxies. Within each redshift interval the galaxies are 
then ordered in terms of apparent magnitude (and therefore crudely in 
absolute magnitude). The progression down the pages qualitatively reflects the 
process of galaxy evolution, although of course does not correct for 
K-corrections and the redshift dependent selection windows.}

\begin{figure}
\epsscale{1.0}
\plotone{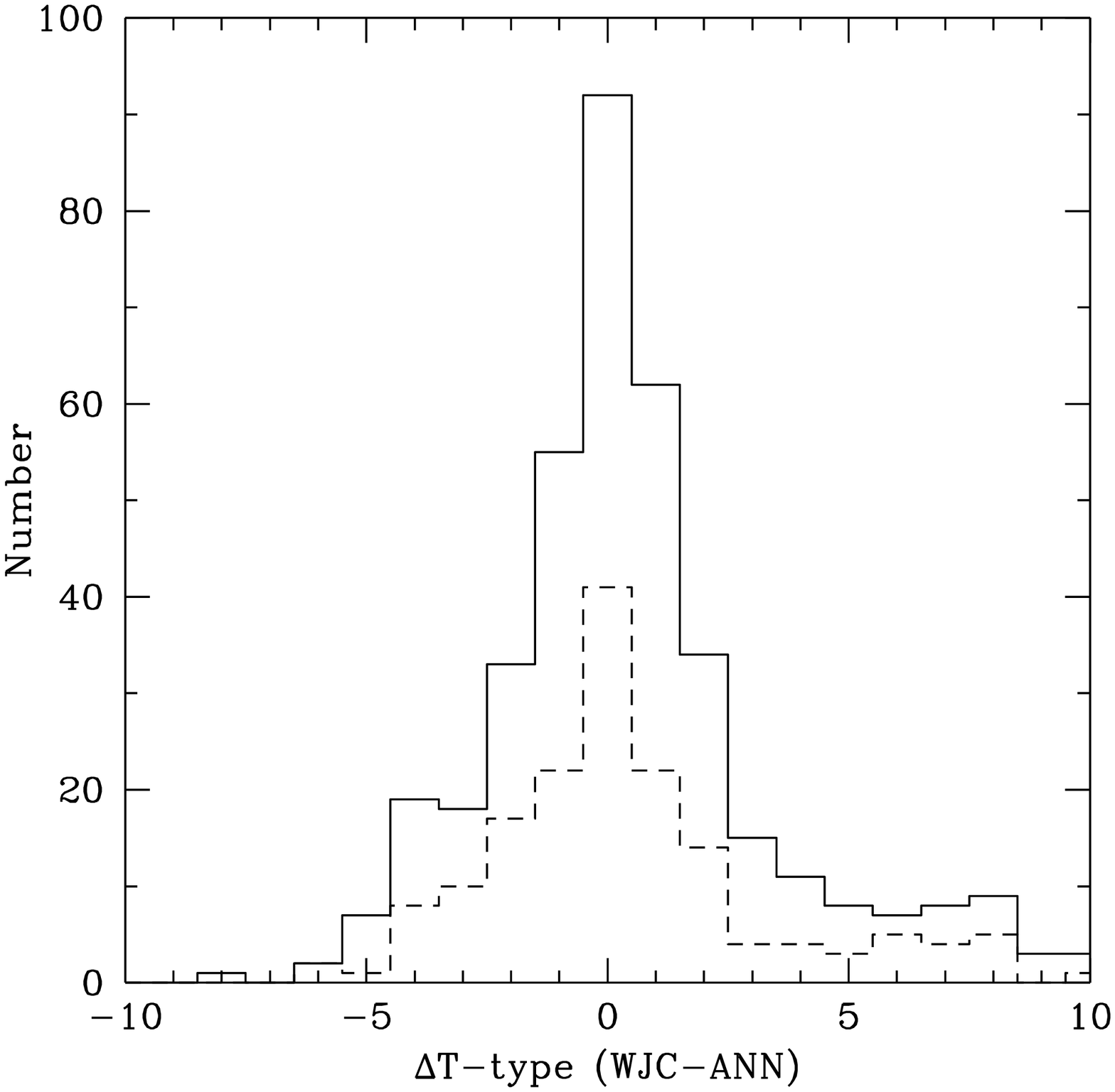}
\end{figure}

\begin{figure}
\epsscale{1.0}
\plotone{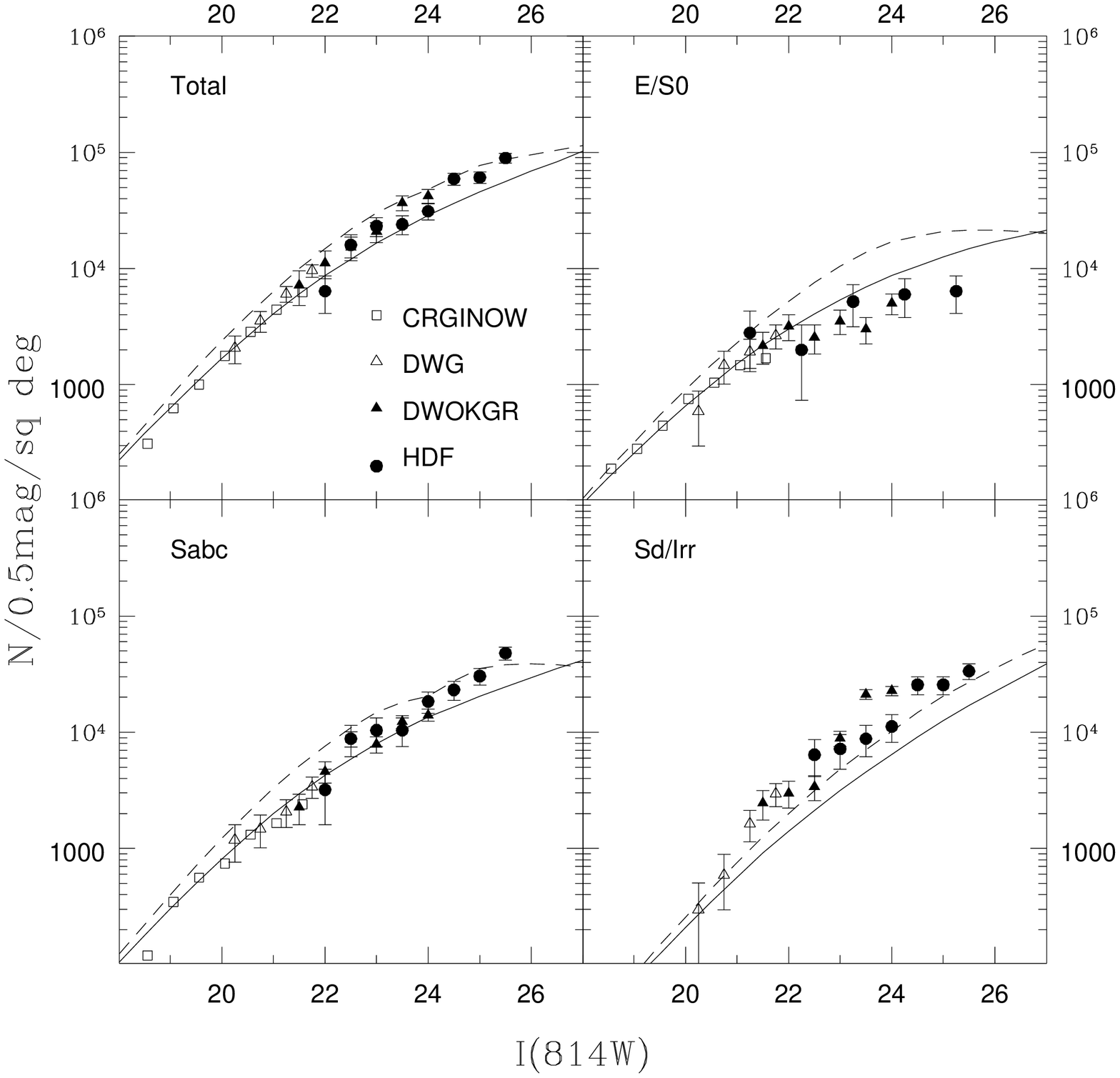}
\end{figure}

\begin{figure}
\epsscale{1.0}
\plotone{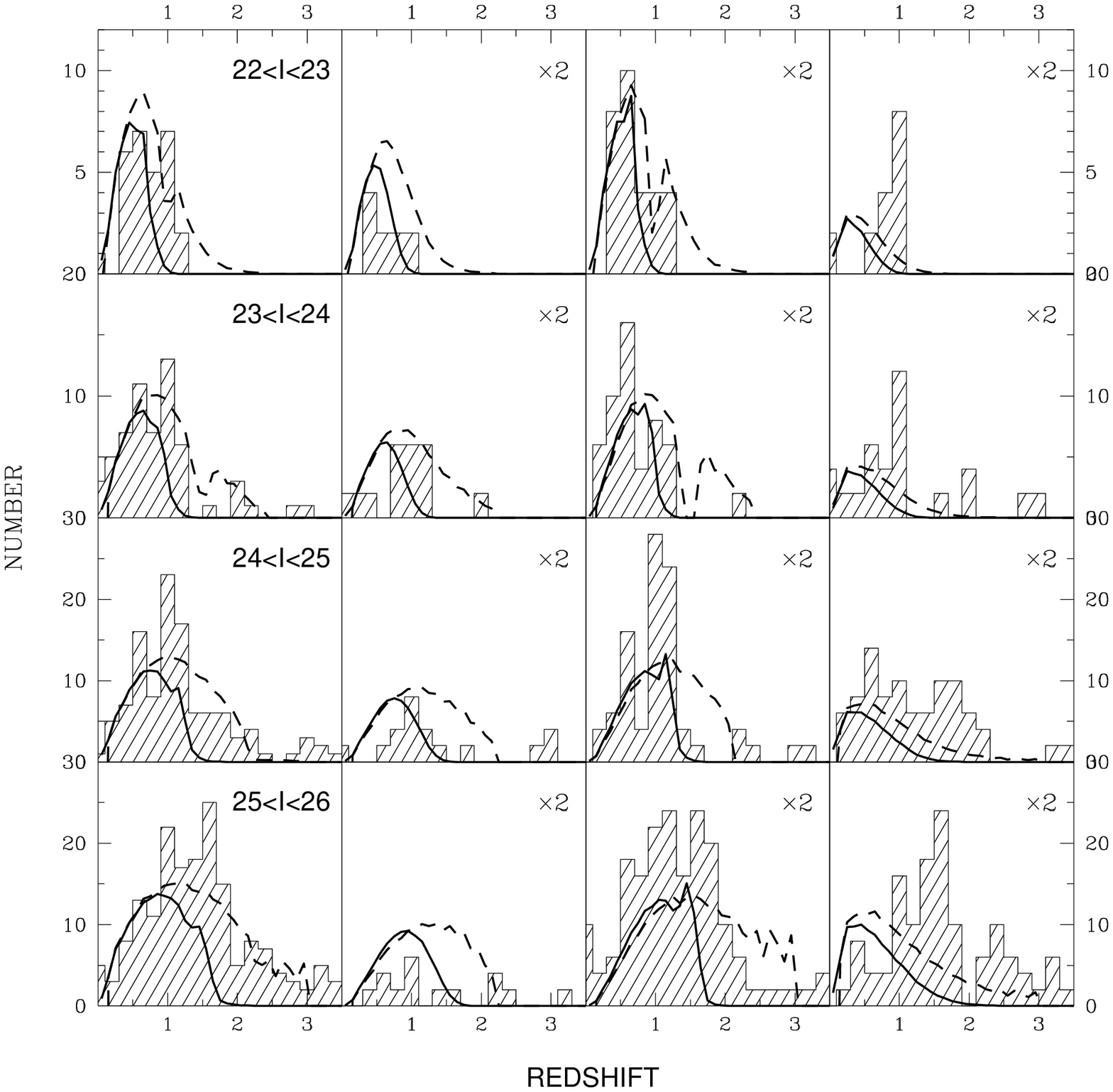}
\end{figure}

\end{document}